\documentclass[a4paper, superscriptaddress, notitlepage, amsmath, amssymb, aip, twocolumn, floatfix, 10pt, jcp]{revtex4-2}
\usepackage{graphicx}
\usepackage{bm,bbm}
\usepackage[urlcolor=blue]{hyperref}
\hypersetup{colorlinks=true,allcolors=blue}
\usepackage{amssymb} 
\usepackage{tcolorbox}

\newcommand{\Elr}[1]{\left\langle #1\right\rangle}
\newcommand{\ee}[1]{{\rm e}^{#1}}
\newcommand{\E}[1]{\langle #1\rangle}
\newcommand{\Es}[1]{\left\langle #1\right\rangle_{\rm s}}
\newcommand{\rmd}{\mathrm{d}}

\newcommand{\f}[1]{\mathbf{#1}}
\newcommand{\x}{x}

\newcommand{\abs}[1]{\left\lvert #1 \right\rvert}

\newcommand{\N}{\mathbb{N}}

\newcommand{\var}{{\rm var}}

\newcommand{\DSi}{\Delta S_{\rm inf}} 
\newcommand{\DSt}{\Delta S_{\rm tot}} 
\newcommand{\DSm}{\Delta S_{\rm med}} 
\newcommand{\DSs}{\Delta S_{\rm sys}} 
\newcommand{\DSb}{\Delta S_{\rm bound}} 

\newcommand{\traj}{(x_\tau)_{0\le\tau\le t}}
\newcommand{\trajBW}{({x}_{t-\tau})_{0\le\tau\le t}}

\newcommand{\trajCG}{(q_\tau)_{0\le\tau\le t}}
\newcommand{\trajCGBW}{(q_{t-\tau})_{0\le\tau\le t}}
\newcommand{\hGamma}{\hat{\Gamma}}
\usepackage{xcolor}
\definecolor{C0}{rgb}{0.12156862745098039, 0.4666666666666667, 0.7058823529411765}
\definecolor{C1}{rgb}{1.0, 0.4980392156862745, 0.054901960784313725}
\definecolor{C2}{rgb}{0.17254901960784313, 0.6274509803921569, 0.17254901960784313}
\definecolor{C2_darker}{rgb}{0.0.14235, 0.517647, 0.0.14235}
\definecolor{C3}{rgb}{0.8392156862745098, 0.15294117647058825, 0.1568627450980392}
\definecolor{C4}{rgb}{0.5803921568627451, 0.403921568627451, 0.7411764705882353}
\definecolor{C5}{rgb}{0.5490196078431373, 0.33725490196078434, 0.29411764705882354}
\definecolor{inferno0}{rgb}{0.087411, 0.044556, 0.224813}
\definecolor{inferno1}{rgb}{0.379001, 0.076253, 0.432719}
\definecolor{inferno2}{rgb}{0.658463, 0.178962, 0.372748}
\definecolor{inferno3}{rgb}{0.894305, 0.353399, 0.193584}
\definecolor{inferno4}{rgb}{0.987622, 0.64532 , 0.039886}

\colorlet{mylinkcolor}{blue!66!black!80}
\hypersetup{colorlinks = true, linkcolor = {mylinkcolor}, linkcolor = {mylinkcolor}, citecolor = {mylinkcolor}, urlcolor = {mylinkcolor}}
\definecolor{grey}{rgb}{0.6,0.6,.6}
\definecolor{darkgrey}{rgb}{0.4,0.4,.4}
\definecolor{darkgreen}{rgb}{0,0.4,0}
\definecolor{lightgreen}{rgb}{0,0.7,0}
\definecolor{darkred}{rgb}{0.5,0,0}

\newcommand{\blue}[1]{{\color{black}#1}}
\newcommand{\oldblue}[1]{{\color{black}#1}}

\newcommand{\dS}{\dot{S}}

\allowdisplaybreaks

\begin{document}
\title{Perspective: 
Time irreversibility in systems observed at coarse
resolution}
\author{Cai Dieball}
\email{cai.dieball@mpinat.mpg.de}
\author{Alja\v{z} Godec}
\email{agodec@mpinat.mpg.de}
\affiliation{Mathematical bioPhysics Group, Max Planck Institute for Multidisciplinary Sciences, 37077 G\"ottingen, Germany}

\begin{abstract}
\oldblue{A broken time-reversal symmetry, i.e.\ broken detailed balance, is central to non-equilibrium physics and is a prerequisite for life. However, it turns out to be quite challenging to unambiguously define and quantify time-reversal symmetry (and violations thereof) in practice, that is, from observations. Measurements on complex systems have a finite resolution and generally probe low-dimensional projections of the underlying
dynamics, which are well known to introduce memory. In situations where many microscopic states become ``lumped'' onto the same observable ``state'' or when introducing 
``reaction coordinates'' to reduce the dimensionality of data, signatures of a broken time-reversal symmetry in the microscopic dynamics become distorted or masked. 
In this perspective we highlight why in
defining and discussing time-reversal symmetry, and quantifying its
violations, the precise underlying assumptions on the microscopic
dynamics, the coarse graining, and further reductions, are \emph{not} a technical detail. These assumptions decide whether the conclusions that are drawn are physically sound
or inconsistent. We summarize recent findings in the field and reflect upon key challenges.}
\end{abstract}
\maketitle

\section{Introduction}
A broken time-reversal symmetry, or broken detailed balance\cite{Gallavotti1995PRL, Kurchan1998JPAMG, Lebowitz1999JSP, Maes1999JSP, Maes2003JSP}, is essential for the existence of
living matter
\cite{Prigogine_1978,Astumian2002Nov,Michaelian_2011}. Stated from a
thermodynamic perspective, the emergence and persistence of life 
requires constant entropy production
\cite{Schrodinger_2012,Michaelian_2011}. These two statements are
intuitive and may nowadays
seem obvious, perhaps rightfully so. However, an attempt to
unambiguously
define time-reversal symmetry (and violations
thereof) and consistently connect it to thermodynamics purely in terms
of \emph{observations} 
turns out to be rather non-trivial. Avoiding the conceptually much
harder problem of the emergence of a positive entropy production
from mechanics (classical or quantum) \cite{Montroll_1954,Zubarev},
the main difficultly we are referring to in this perspective is to define, understand, and
infer violations of detailed balance in a system's dynamics when one
cannot observe all degrees of freedom. There may be cases, where one at least
has some knowledge about how the \emph{observable} depends on
the relevant (i.e., ``slow''; to be made precise later),  hidden microscopic degrees of freedom, and in some cases
one does not even have such knowledge.

From a practical perspective, measurements on complex systems either have a finite
resolution~\cite{Gladrow2016PRL,Battle2016S,Dieball2022PRL,Dieball2022PRR} or do
\emph{not} resolve all relevant degrees of
freedom---they only probe low-dimensional projections of the
dynamics, such as the magnetization \cite{Ocio} or dielectric
response \cite{dielectric}, and observables such as
molecular extensions or \blue{F\"orster resonance energy transfer} efficiencies and lifetimes in
single-molecule experiments \cite{order,Woly,Xie,Dima_2013,Brujic,Hagen,Dima_test,Lapolla2021PRR}. It
is well known that these projections typically introduce memory
\cite{Mori1965PTP,Zwanzig1973JSP,Lapolla2019FP}. In
these situations, many microscopic states become ``lumped'' onto the same
observable ``state''. Similar projections emerge when introducing 
``reaction coordinates'' to reduce the dimensionality of data
to the essential observable(s)
\cite{Weinan_2004,Peters_2016,Banushkina_2016}, which is taken a step
further if one constructs kinetic (typically Markov) models with a small number of most relevant,
long-lived states \cite{Prinz_2011,Schwantes_2014,No__2019}. These
projections and reductions significantly distort or mask signatures of a broken time-reversal symmetry in microscopic
dynamics. 
The main aim of this perspective is \emph{not} on biological applications \cite{Gnesotto_2018}, but to highlight why in
defining and discussing time-reversal symmetry, and quantifying its
violations, the precise underlying assumptions on the microscopic
dynamics, the projection/coarse graining, and further reductions, are \emph{not} 
merely a technical detail. These assumptions typically decide whether the
conclusions that are drawn are physically correct
or inconsistent, and in turn whether they lead to controversy.      

Defining the entropy production rate based on some general
observable (i.e., without knowledge about microscopic dynamics) from
first principles is non-trivial. While irreversibility implies that \emph{microscopic} trajectories evolving ``forwards'' in time are not equally
likely as their time-reversed counterparts, the situation is
typically very different when we observe only projections and/or discretized
trajectories. To motivate this claim, we note the ``oddness'' of certain degrees
of freedom (like momenta) under time-reversal. Note that the
sign-change is a \emph{post factum} rule, there is no necessary mathematical
condition for assuming it, it is enforced to align with observations.

Inertia is essentially nothing but the
persistence of motion, and memory effects may, consistently, be considered a
generalization of this (trivial) persistence. In other words, memory
effects are, strictly speaking, manifested as (anti-)persistence of
motion---a particle is more (or less) likely to move in the
instantaneous direction or to transition into another state depending
on the past. It should not be difficult to appreciate that deciding
about a possible ``sign change'' of general memory effects is far from trivial. In
other words, in general (except if assuming that everything that is
hidden is at equilibrium \cite{Aron2010JSMTE}) one cannot simply ``flip the sign'' of memory effects.    

This perspective is written to highlight, by examples, possible
inconsistencies of, and incompatibilities between, common underlying
assumptions. For the sake of simplicity we will focus on situations
where the ``microscopic'' dynamics is an \emph{overdamped  Markovian
dynamics}. To ease the reading, we refrain from being rigorous, but
we nevertheless insist on being consistent and, as far as feasible,
precise. We will express entropy in units of Boltzmann's constant
$k_{\rm B}$ and energy in units of thermal energy $k_{\rm B}T$ at temperature $T$.  

\section{Entropy production in Markovian systems coupled to an
  equilibrium bath}
A meaningful definition of entropy production ought to measure the
discrepancy between the probability of forward paths,
$(x_\tau)_{0\le\tau\le t}$, $\mathbbm{P}[(x_\tau)_{0\le\tau\le t}]$,
and the probability of time-reversed paths,
$(\theta x_\tau)_{0\le\tau\le t}$, $\widetilde{\mathbbm{P}}[(\theta x_\tau)_{0\le\tau\le t}]$\oldblue{, where the backwards paths may be measured with a different probability measure $\widetilde{\mathbbm{P}}$ than the forwards paths, and hence may refer to a different ensemble \cite{Seifert2018PA}}. A priori there are many possibilities to quantify a discrepancy, 
e.g., any $L^p$ norm. However, the definition for Markovian dynamics that turns out to be
consistent with classical (phenomenological) thermodynamics is the
Kullback-Leibler divergence \cite{Seifert2012RPP,Seifert2018PA},
\begin{align}
S_{\rm inf}
&=\int \rmd\mathbbm{P}[(x_\tau)_{0\le\tau\le t}]\ln\frac{\mathbbm{P}[(x_\tau)_{0\le\tau\le t}]}{\widetilde{\mathbbm{P}}[(\theta x_\tau)_{0\le\tau\le t}]}.
\label{EPR}
\end{align}
\blue{Note that for the Kullback-Leibler divergence to exist one mathematically requires that $\mathbb{P}$ is absolutely continuous with respect to $\widetilde{\mathbbm{P}}$ which is from a physical point of view identified as the local detailed balance condition.}
 $S_{\rm inf}\equiv S_{\rm inf}[0,t]$ is called the \emph{informatic} entropy production in the
interval $[0,t]$ \cite{Seifert2018PA}, as it a priori does not
necessarily relate to the thermodynamic entropy production.
\oldblue{One may also write $S_{\rm inf}=\langle \Delta s_{\rm inf}[(x_\tau)_{0\le\tau\le t}]\rangle$} which is the expectation $\E{\cdots}$ of the stochastic path-wise entropy
production
$\Delta s_{\rm inf}[(x_\tau)_{0\le\tau\le t}]\equiv \ln(\blue{\mathbbm{P}[(x_\tau)_{0\le\tau\le t}]}/\widetilde{\mathbbm{P}}[(\theta
      x_\tau)_{0\le\tau\le t}])$
  \cite{Seifert2005PRL} in the forward
path ensemble $\mathbbm{P}[(x_\tau)_{0\le\tau\le t}]$. Only for
a physically consistent choice of time reversal $\theta$ \oldblue{and backwards measure $\widetilde{\mathbbm{P}}$},  
\blue{the informatic and thermodynamic entropy production} 
may become equivalent \cite{Seifert2005PRL,Maes2008MPRF,MuratoreGinanneschi2013JSP,Chetrite2008CMP}, which will be discussed in more detail below. 

From $\frac{\widetilde{\mathbbm{P}}[(\theta
  x_\tau)_{0\le\tau\le t}]}{\mathbbm{P}[(x_\tau)_{0\le\tau\le t}]}=\ee{-\Delta s_{\rm
    inf}[(x_\tau)_{0\le\tau\le t}]}$ and
\begin{align*}
  1&=\int {\rm d}\mathbbm{P}[(x_\tau)_{0\le\tau\le t}]
  =\int \rmd\mathbbm{P}[(x_\tau)_{0\le\tau\le t}]\frac{\widetilde{\mathbbm{P}}[(\theta
  x_\tau)_{0\le\tau\le t}]}{\mathbbm{P}[(x_\tau)_{0\le\tau\le t}]}
      \\&=\Elr{\ee{-\Delta s_{\rm
    inf}[(x_\tau)_{0\le\tau\le t}]}}
    \ge \ee{-\Delta S_{\rm
    inf}}, 
\end{align*}
where in the last step we used Jensen's inequality for the convex
exponential function and Eq.~\eqref{EPR}, 
follows the detailed fluctuation theorem $\E{\ee{-\blue{\Delta s_{\rm
      inf}}}}=1$ and non-negativity of the entropy production (i.e., the
second law) $\Delta
S_{\rm inf}\ge 0$.

For brevity and sake of simplicity we will limit the subsequent
discussion to situations, where the detailed, ``microscopic'' dynamics corresponds
to an overdamped diffusion (with additive noise) in $d$ dimensions and obeys local detailed balance\cite{Seifert2012RPP,Falasco2021PRE,Maes2021SPLN}. Then, the relevant (``slow'') degrees of freedom
$x_t\in\mathbbm{R}^d$ obey the It\^o equation  with symmetric positive-definite
$d\times d$ diffusion matrix $D$ proportional to the temperature, $D\propto T$ reflecting the effect of a heat bath containing all irrelevant (``fast'') degrees of freedom (at all times) at equilibrium at temperature $T$, and a drift $a(x):\mathbbm{R}^d\to\mathbbm{R}^d$ for simplicity without explicit time dependence \cite{Gardiner1985},
\begin{align}
  \rmd x_\tau= a(x_\tau)\rmd\tau+\sqrt{2D}\rmd W_\tau,
  \label{SDE}
\end{align}
with $\E{\rmd W^i_\tau}=0$ and $\E{\rmd W^i_\tau\rmd
  W^j_{\tau'}}=\delta_{ij}\delta(\tau-\tau')\rmd\tau\rmd\tau'$. 
The probability density of $x_\tau$, $p(\x,\tau)$, obeys the
Fokker-Planck equation \cite{Risken1989}, which is a continuity
equation  $\partial_t p(x,\tau)=-\nabla\cdot j(x,\tau)$ with
probability current $j(x,\tau)\equiv \nu(x,\tau)p(x,\tau)$ that depends on the
\emph{local mean velocity} $\nu(x,\tau):\mathbbm{R}^d\times
     [0,\infty)\to \mathbbm{R}^d$, defined as
\begin{align}
\nu(x,\tau)&\equiv\frac{\left[a(x)-D\nabla \right]p(x,\tau)}{p(x,\tau)}.\label{FPE}
\end{align}

\subsection{Thermodynamic entropy production}
While path-wise, stochastic definitions of the entropy production exist \cite{Seifert2005PRL}, throughout this perspective we focus on the deterministic, average values. 
We first address steady-state dynamics \cite{Jiang2004} with invariant density $p_{\rm
  s}(x)$, i.e., we first consider strongly ergodic systems relaxing to
an invariant density
$p(x,\tau\to\infty)=p_{\rm s}(x)$, which requires $a(x)$ to be
sufficiently confining. We either prepare the system in the steady
state, i.e., $x_{\tau=0}\sim p_{\rm s}(x)$, or begin our observations
after a sufficiently long time.  
Such systems may be in thermodynamic equilibrium (these are referred
to as being in ``detailed balance''), where no entropy is
produced. Conversely, a system in the stationary state may also
perform work $W$ against the friction \cite{Seifert2012RPP}
\begin{align}
W
&=\Es{\int_0^t [D^{-1}a(x_\tau)]\cdot\circ\rmd
x_\tau}\nonumber\\
&=t \Es{\nu_{\rm s}^T(x)D^{-1}\nu_{\rm s}(x)}\ge 0,
\label{W}
\end{align}
where $\Es{\cdots}$ denotes the steady-state ensemble
average, $\circ$ the stochastic Stratonovich integral, and we
introduced  the invariant local mean velocity $\nu_{\rm
  s}(x)=\nu(x,\tau\to\infty)$. The non-negativity of $W$ follows from
the fact that $D$ is positive definite. Steady-state systems with $W>0$ are said to be out of equilibrium or
dissipative.  The system is connected to a heat bath with temperature
$T$, such that the work $W$ gives rise to an entropy production in the
medium $\DSm=W/T$, where the $\Delta$ notation always refers to
entropy produced in the time interval $[0,t]$, i.e., $\Delta S\equiv S(t)-S(0)$. In
steady states \oldblue{under the assumptions on the heat bath mentioned above}, this is the only source of entropy production, such
that the total thermodynamic entropy production is given by \cite{Seifert2012RPP}
\begin{align}
  \DSt=\DSm\ge0\,.
  \label{EPRs}
\end{align}
For any non-stationary preparation of the system, $p(x,\tau=0)\ne p_{\rm
  s}(x)$, the probability density changes in time and there is an
additional contribution to entropy production, namely the change of
Gibbs-Shannon entropy in the system  \cite{Seifert2012RPP}
\begin{align}
    \DSs =  - \int\rmd x\,p(x,t)\ln p(x,t) +
    \int\rmd x\,p(x,0)\ln p(x,0)\,,
\end{align}
which can also take on negative values. Similarly, $\DSm$ (and thus $W$) can be
positive or negative for non-steady-state dynamics. During transients
the total thermodynamic entropy production reads \cite{Seifert2005PRL,Seifert2012RPP},
\begin{align}
  \DSt &= \DSm + \DSs\nonumber\\
  &= \int_0^t\rmd\tau\langle \nu^T(x,\tau)D^{-1}\nu(x,\tau)\rangle
  \ge 0\,,
  \label{Stot SDE}
\end{align}
where $\langle\cdot\rangle$ denotes the expectation over
$p(x,\tau)$. At thermodynamic equilibrium $\DSt=0$.

Although $\DSs$ and $\DSm$ can be positive, negative, or zero, we note
that it is \emph{not} possible to have $\DSs=-\DSm\ne0$. That is, we
cannot have $\DSt=0$ when $\DSs\ne0$ or $\DSm\ne0$. To see this, note
that $\DSs\ne0$ implies that $\partial_\tau p(x,\tau)\ne 0$, which via
the Fokker-Planck equation \eqref{FPE} gives $\nu(x,\tau)\ne 0$ for
some $x,\tau$, in turn implying $\DSt\ne0$ via Eq.~\eqref{Stot
  SDE}. The fact that any evolution of $p(x,\tau)$ bounds $\DSt$ from
below by a positive constant is formally encoded in the
Benamou-Brenier formula \cite{Benamou2000NM,VanVu2023PRX} and further
in thermodynamic bounds on transport \cite{Dieball2024PRL}.

\subsection{Equivalence of thermodynamic and informatic entropy
  production for overdamped Markov dynamics}\label{equival} 

Whereas the Onsager-Machlup formula for the ``stochastic action''
entering path probabilities \cite{Onsager1953PR} in Eq.~\eqref{EPR} are well known in the
physics literature, their meaning is far from trivial. Namely, the
action of a path $(x_\tau)_{0\le\tau\le t}$ governed by
Eq.~\eqref{SDE} is assumed to read 
\begin{align}
&A[(x_\tau)_{0\le\tau\le t}]\label{action}
\\
&=\frac{1}{4}\int_0^t\rmd \tau\{[\dot{x}_\tau-a(x_\tau)]^TD^{-1}[\dot{x}_\tau-a(x_\tau)]+2\nabla\cdot a(x_\tau)\},\nonumber
\end{align}  
\blue{where $\dot{x}_\tau\equiv\frac{\rmd x_\tau}{\rm d \tau}$ and note that
the last term does \emph{not} appear in the small-noise limit \cite{Freidlin_2012}
but is necessary in general (see Theorem 9.1 and Corollary in Chapter VI of Ref.~\cite{Ikeda1981}). Although it is often claimed in the literature that the last term only appears when interpreting the dynamics in the Stratonovich convention (see, e.g., Ref.~\cite{Lau2007PRE}), this appears to be a pure misconception. First, the choice of convention is fixed when stating the equation of motion Eq.~\eqref{SDE}. Second, for additive noise ($D$ independent of $x$) the convention  does \emph{not} influence the statistics of paths $\traj$ and consequently the action $A$ in Eq.~\eqref{action} is also the same. In the Appendix, we summarize different strategies to derive Eq.~\eqref{action}, and highlight why for all strategies the only correct form of this equation has to include $\int_0^t\rmd \tau\nabla\cdot a(x_\tau)/2$.}

Notably, it
is well known that $\mathbbm{P}(|\dot{x}_\tau|<C)=0$ for any $C<\infty$ and any $\tau$,
so the $\dot x_\tau$ in Eq.~\eqref{action} should cause (substantial) concern,
since paths are locally almost surely non-differentiable. 
The issue
may be resolved by recalling the \emph{support theorem of diffusion} 
\cite{Stroock_1972}, according to which (loosely speaking) the probability measure of nowhere
differentiable diffusion paths
$(x_\tau)_{0\le\tau\le t}$ has support on the closure of smooth paths
$(y(\tau))_{0\le\tau\le t}\in C^\infty$,
as long as said smooth approximations are interpreted according to the
Stratonovich convention. 
The Stratonovich formulation is required for consistency between
systems with ``smooth'' and ``Brownian'' inputs (see also
\cite{WongZakai}). In this context, we ought to make the
replacement $x_\tau\to y(t)$ in Eq.~\eqref{action} along the following
lines.

The rigorous probabilistic definition of the Onsager-Machlup
functional (see Theorem 9.1 and Corollary in Chapter VI of Ref.~\cite{Ikeda1981}) follows from a comparison of probabilities of $x_\tau$
remaining within a ``tube'' of radius $\varepsilon$ around a smooth
path $y(t)\in C^\infty$ relative to the probability that the Wiener
process remains within a straight ``tube'' of radius $\varepsilon$. To
state this precisely we note that 
\begin{align}
{\mathbbm{P}}[(x_\tau)_{0\le\tau\le t}]&=\mathbbm{P}[(x_\tau)_{0\le\tau\le t} | x_0]p(x_0,0)\nonumber \\
\widetilde{\mathbbm{P}}[(\theta
  x_\tau)_{0\le\tau\le t}]&=\mathbbm{P}[(\theta
  x_\tau)_{0\le\tau\le t} | x_t]p(x_t,t),
  \label{conditioned}
  \end{align}
and as well as the supremum norm
\begin{align}
\|y(\tau)-x_\tau\|_{0\le \tau\le t}\equiv\sup_{0\le\tau\le t}\sup_{i=1,\dots,d}|y_i(\tau)-x_i(\tau)|,
\label{norm}  
\end{align}
such that 
\begin{align}
\lim_{\varepsilon\to
  0}\frac{\mathbbm{P}\left[\|y(\tau)-x_\tau\|_{0\le\tau\le
      t}<\varepsilon\right]}{\mathbbm{P}[\|\sqrt{2D}W_\tau\|_{0\le\tau\le t}<\varepsilon]}=\mathrm{e}^{-A[(y(\tau))_{0\le\tau\le t}]}p(y(0),0),
\label{OM}  
\end{align}
where $p(y(0))$ is the probability density of the initial
condition and we used Eq.~\eqref{conditioned}. The same holds true for the time-reversed path, which for
overdamped Markov processes is simply $(\theta x_\tau)_{0\le\tau\le t}=(x_{t-\tau})_{0\le\tau\le t}$, which using Eq.~\eqref{conditioned} reads \cite{Seifert2018PA}
\begin{align}
\lim_{\varepsilon\to 0}&\frac{\mathbbm{P}\left[\|y(t-\tau)-x_{t-\tau}\|_{0\le\tau\le
      t}<\varepsilon\right]}{\mathbbm{P}[\|\sqrt{2D}W_\tau\|_{0\le\tau\le t}<\varepsilon]}
 \nonumber     \\&=\mathrm{e}^{-A^\dagger[(y(\tau))_{0\le\tau\le t}]}p(y(t),t),
\label{OMb}  
\end{align}
where we introduced the ``backward action''
\begin{align}
A^\dagger[(y(\tau)&)_{0\le\tau\le t}]=\frac{1}{4}\int_0^t\rmd \tau\{
[\dot{y}(\tau)+a(y(\tau))]^T\times
\nonumber\\& D^{-1}[\dot{y}(\tau)+a(y(\tau))]+2\nabla\cdot a(y(\tau))\}.\label{b_action}
\end{align} 
Note that thermodynamic consistency requires that the probability
measures of forward and backwards paths are equivalent, i.e., $\mathbbm{P}[(x_\tau)_{0\le\tau\le t}]>0\iff\widetilde{\mathbbm{P}}[(\theta x_\tau)_{0\le\tau\le t}]>0$ \cite{Seifert2012RPP}. The
fraction in Eq.~\eqref{EPR} is non-zero on the support of
$\mathbbm{P}[(x_\tau)_{0\le\tau\le t}]$. As a result we have on the
support of $\mathbbm{P}[(x_\tau)_{0\le\tau\le
    T}]$
\begin{align}
\frac{\mathbbm{P}[(y(\tau))_{0\le\tau\le t}]}{\mathbbm{P}[(y(t-\tau))_{0\le\tau\le t}]}
&=\frac{\lim_{\varepsilon\to
  0}\frac{\mathbbm{P}\left[\|y(\tau)-x_\tau\|_{0\le\tau\le
      t}<\varepsilon\right]}{\mathbbm{P}[\|\sqrt{2D}W_\tau\|_{0\le\tau\le t}<\varepsilon]}}{\lim_{\varepsilon\to
  0}\frac{\mathbbm{P}\left[\|y(t-\tau)-x_{t-\tau}\|_{0\le\tau\le
      t}<\varepsilon\right]}{\mathbbm{P}[\|\sqrt{2D}W_\tau\|_{0\le\tau\le t}<\varepsilon]}}
     \nonumber \\&=\mathrm{e}^{-A+A^\dagger}\frac{p(y(0),0)}{p(y(t),t)}.
\label{coomute}
\end{align}
and in turn, since
$-A+A^\dagger=\int_0^ta(y(\tau))^TD^{-1}\dot{y}(\tau)\rmd \tau=\int_0^t[D^{-1}a(y(\tau))]\cdot\dot{y}(\tau)\rmd \tau$, we have on the
support of $\mathbbm{P}[(x_\tau)_{0\le\tau\le t}]$
\begin{align}
\Delta s_{\rm inf}[(x_\tau)_{0\le\tau\le t}]=\int_{\tau=0}^{\tau=t}
[D^{-1}a(x_\tau)]\cdot\circ\rmd
x_\tau+\ln\frac{p(x_0,0)}{p(x_t,t)}.
\label{s_inf_1}
\end{align}
For steady-state dynamics we have $p(x,0)=p(x,t)=p_{\rm
  s}(x)$ such that the
logarithmic term vanishes, and upon taking the average and using Eq.~\eqref{FPE}
we recover $\Delta S_{\rm inf}=W/T$ with $W$ given in Eq.~\eqref{W}. More generally, for any initial
condition $p(x_0,0)$ we obtain Eq.~\eqref{Stot SDE}. For overdamped
diffusions the informatic and thermodynamic entropy production are
hence equivalent, $\DSt =\Delta S_{\rm
    inf}$. Note that these results follow more directly
from the Radon-Nikodym theorem \cite{Min_1999}.

Another deep path-wise result we want to mention here is known as
Watanabe's formula (see Refs.~\cite{Min_1999,Qian2021PRE} and Theorem 9.2
in Ref.~\cite{Ikeda1981}) which states for overdamped diffusion that any smooth closed curve
$\Gamma\in C^\infty$, i.e., $\bar{y}(\tau)\in C^\infty$ in $\mathbbm{R}^d$ such
that $\bar{y}(0)=\bar{y}(t)$, we have
\begin{align}
\frac{\mathbbm{P}[(\bar{y}(\tau))_{0\le\tau\le
      t}|y(0)]}{\mathbbm{P}[(\bar{y}(t-\tau))_{0\le\tau\le
      t}|y(t)]}
      &=\exp\left(\int_0^ta(\bar{y}(\tau))^TD^{-1}\dot{\bar{y}}(\tau)\rmd
\tau\right)
\nonumber\\&=\exp\left(\oint_\Gamma[D^{-1}a(\bar y)]\cdot \rmd \bar{y}\right),
\end{align}  
where $\oint_\Gamma D^{-1}a\cdot \rmd \bar{y}$ is known as the ``cycle
force'' \cite{Qian_cycle}. The fraction equals 1 if and only if $x_t$ obeys detailed
balance, $ D^{-1}a=-\nabla\phi$ (i.e., when $D^{-1}a(x)$ is an exact
one-form).

\section{Dynamics at coarse resolution}

\subsection{Markovian coarse-grained dynamics}\label{MarkovJ}

\begin{figure}
    \centering
    \includegraphics[width=.49\textwidth]{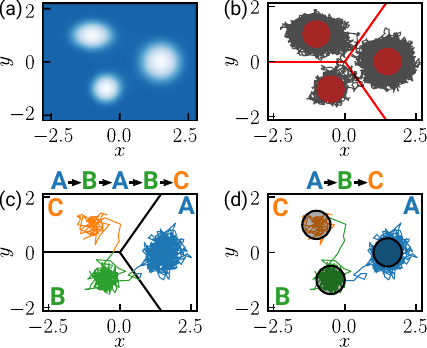}
    \caption{(a) A free-energy landscape $\phi(\f x)$ (white to blue) with three minima. (b) Overdamped motion $\rmd\f x_\tau = -D\nabla \phi(\f x_\tau)+\sqrt{2D}\rmd\f W_\tau$ (grey) can be coarse grained into three states using lumping (e.g., see three areas separated by lines) or using three milestones (e.g., see circles). Only the latter procedure will approach Markovian dynamics for deep and well separated minima. (c) Lumping and (d) milestoning sketched for part of the trajectory with a reduced time resolution.}
    \label{fig:milestone_2d}
\end{figure}

\begin{figure*}
\includegraphics[width=\textwidth]{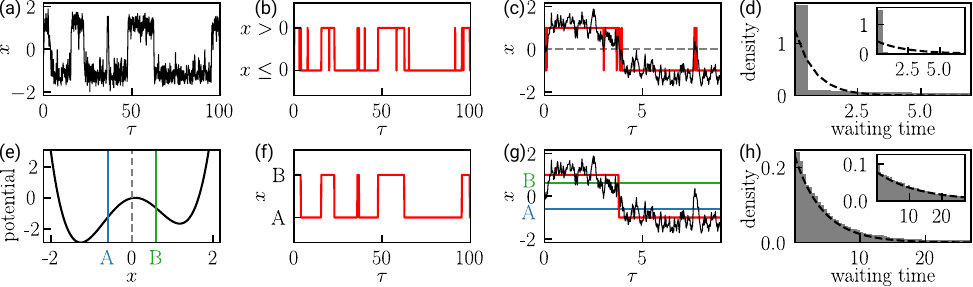}
\caption{Numerical evaluation of one-dimensional barrier crossing. (a,
  e) \oldblue{Dynamics $\rmd
    x_\tau=-\partial_x\phi(x_\tau)\rmd\tau+\sqrt{2}\rmd W_\tau$ evolving in
    a double-well potential $\phi(x)=x^4 - 3x^2 + x/2$ simulated with
    $\rmd\tau=0.01$}. The sample trajectory (a) is coarse-grained into
  two states by lumping $x\le 0$ and $x>0$ (b) or milestoning
  \oldblue{with $A=-0.6$ and $B=0.6$} (f), where in the latter procedure
  the coarse-grained state changes once the milestones $A$ or $B$ are
  reached. Zoom into the initial part of the trajectory highlights
  frequent state changes (``fast re-crossings'') in the lumped
  dynamics (c) compared to (g) the milestoned
  dynamics. Waiting time density
  for state changes inferred as a histogram from a very long
  trajectory \oldblue{($5\times 10^{7}$ time steps)} for  (d) lumped for
  the transition $x>0
  \to x\le 0$ (Inset: for the transition $x\le 0\to x>0$)  and (h)
  milestoned  for
  the transition $B\to A$ (Inset: for the transition $A\to B$) dynamics reveal that only the latter agrees with the
  desired exponential (black dashed line is $\lambda\exp(-\lambda t)$
  for $\lambda^{-1}=$ mean waiting time). \blue{Note that the lumped
    trajectories (b, c) 
    explicitly depend on the
    time resolution and do not converge as $\rmd\tau\to 0$ (more precisely, in the mathematical idealization $\rmd
    x_\tau=-\partial_x\phi(x_\tau)\rmd\tau+\sqrt{2}\rmd W_\tau$, there will be a diverging number of recrossings such that the waiting time density (d) for a fixed trajectory of finite length $t$ converges to a delta function at $0$).}}
\label{fig:Kramers} 
\end{figure*}

When the (generalized) potential $\phi(x)\equiv-\ln p_{\rm s}(x)$
features deep wells separated by high (and sharp) barriers
\cite{Moro_1995} and under
some additional (but commonly met) requirements on the irreversible
component of $a(x)$, $a(x)-D\nabla\phi(x)$ \cite{Maier1993PRE}, the
time-scales of so-called ``librational'' motions inside the wells and transitions
between the wells are sufficiently separated. Moreover, if (and only if
\cite{Moro_1995,Hartich2021PRX,Suarez2021JCTC,Hartich2023PRR}) the
dynamics are projected onto sufficiently localized cores centered at
the minima \cite{Moro_1995} (also referred to as ``milestones''
\cite{Schuette2011JCP,Hartich2021PRX,Suarez2021JCTC}), \oldblue{(see Fig.~\ref{fig:milestone_2d} as well as the Kramers-type result in Fig.~\ref{fig:Kramers})}, the resulting slow,
long-time-scale ``hopping'' dynamics between the cores is a Markov process. That is, time-scale
separation ensured by the nominal features of $a(x)$ is by itself
\emph{not} sufficient for the emerging process to be Markov, a
particular type of coarse graining, nowadays commonly referred to as
``milestoning''
\cite{Elber_2004,Schuette2011JCP,Suarez2021JCTC,Hartich2021PRX,Blom2024PNASU} is
required as well. The reason why naively ``cutting'' the configuration
space into pieces---a process referred to as ``lumping''---cannot give
rise to Markov dynamics on the discrete state space is that by lumping we
(i) inherently (and inevitably) include the fast librational motions
and rapid
re-crossings of the barriers (see Figs.~\ref{fig:milestone_2d}c and \ref{fig:Kramers}c) and (ii) the transition probabilities and
waiting times in the coarse-grained states depend on previous states
\cite{Moro_1995,Hartich2021PRX,Hartich2023PRR}. Note that issues (i)
and (ii) equally emerge when a many-state discrete-state jump process
is lumped to fewer states.

Recall that as soon as
the waiting time in the discrete states is not exponentially
distributed or depends on previous states, the coarse-grained jump
process is \emph{not} Markovian. To make this explicit, only
milestoning-type of coarse grainings may yield Markov jump dynamics
in the presence of a time-scale separation, whereas lumping does
\emph{not} \cite{Schuette2011JCP,Hartich2021PRX,Suarez2021JCTC}. Beside not calling it
``milestoning'', this fact was realized already by Kramers in his
seminal work \cite{Kramers1940P}\blue{. The Kramers result is displayed in Fig.~\ref{fig:Kramers} where the waiting-time density of a two-state system is shown after lumping (see Fig.~\ref{fig:Kramers}b-d; far from the  exponential line) and milestoning (see Fig.~\ref{fig:Kramers}f-h; approaching the exponential in the limit of a very high and sharp barriers and well-separated milestones)}. 
It is also very well established in the community
building Markov-state models \cite{Schuette2011JCP,Suarez2021JCTC}. Yet, it has only
recently becoming appreciated in Stochastic Thermodynamics
\cite{Hartich2021PRX,Hartich2023PRR,Blom2024PNASU,vanderMeerPRX2022,vanderMeer2023PRL,Degunther2024PNASU}
which we will address below in more detail.

Assuming the coarse-grained dynamics (to a sufficient approximation)
corresponds to a Markov process
(i.e., implicitly assuming that it emerged from a milestoning-type of
coarse-graining\oldblue{, see Fig.~\ref{fig:milestone_2d}d}),  a trajectory $(x_\tau)_{0\le\tau\le t}$ corresponds
to a sequence of digitized, say $N$, states. The evolution of the
probability of said states encoded in the probability vector
$p(t)\in(\mathbbm{R}^+)^N$ follows a master equation
\cite{Schnakenberg1976RMP,Gardiner1985} $\frac{\rmd}{\rmd t}p(t)=Lp(t)$
where $L$ is an $N\times N$ matrix whose elements $L_{ij}=r_{ji}$ for $i\ne j$ are
transition rates between states $j\to i$ and $L_{ii}=-\sum_{j\ne
  i}L_{ji}$. We remind
ourselves that the Markov-jump process is an approximation to the long-time-scale
dynamics of the microscopic system only, i.e., temporal resolution is lost. Therefore,
Markov-jump models correspond to a \emph{less general} class of
dynamics and are in fact already captured (including 
short-time motions) in diffusion models, a fact that is occasionally forgotten. 

Analogously to overdamped diffusion in Eq.~\eqref{W}, a system in a
non-equilibrium stationary state with state probability vector $p_{\rm s}$ may
perform work, which in the
Markov-jump case reads \cite{Seifert2012RPP}
$W=\frac12\sum_{i,j}[(p_{\rm s})_jr_{ji}-(p_{\rm s})_ir_{ij}]\ln({r_{ji}}/{r_{ij}})$, giving rise to an entropy
production in the medium $\DSm=W/T$. In steady states, this is
the only source of entropy production, such that $\DSt=\DSm\ge0$. For any non-stationary preparation of the system, $p(0)\ne p_{\rm
  s}$, the probability density changes in time and there is an
additional contribution to entropy production---the change of
Gibbs-Shannon entropy in the system
\cite{Schnakenberg1976RMP,Seifert2012RPP,Peliti2021},
\begin{align}
    \DSt =&\int_0^t\rmd\tau\bigg[\frac12\sum_{i,j}[p_j(\tau)r_{ji}-p_i(\tau)r_{ij}]\ln\frac{r_{ji}}{r_{ij}} 
    \nonumber\\&+ \frac12\sum_{i,j}[p_j(\tau)r_{ji}-p_i(\tau)r_{ij}]\ln\frac{p_j(\tau)}{p_i(\tau)}\bigg]\label{Stot MJP}\\
    =&\frac12\int_0^t\rmd\tau\sum_{i,j}[p_j(\tau)r_{ji}-p_i(\tau)r_{ij}]\ln\frac{p_j(\tau)r_{ji}}{p_i(\tau)r_{ij}}\ge 0\,,\nonumber
\end{align}
where non-negativity follows immediately from the fact that
$p_j(\tau)r_{ji}-p_i(\tau)r_{ij}$ and
$\ln[{p_j(\tau)r_{ji}}/{p_i(\tau)r_{ij}}]$ always have the same sign.
\oldblue{Similar to Sec.~\ref{equival}, also for Markov-jump processes the correct time reversal that yields $\DSt=\DSi$ is $(\theta
x_\tau)_{0\le\tau\le t}=(x_{t-\tau})_{0\le\tau\le t}$ \cite{Seifert2012RPP,Seifert2018PA}.}
Essential for thermodynamic consistency of Markov-jump
processes is the so-called \emph{local detailed balance}
\cite{Seifert2012RPP,Falasco2021PRE,Maes2021SPLN,Hartich2021PRX,Hartich2023PRR}, which in this particular
case corresponds to
\begin{align}
\ln\frac{r_{ij}}{r_{ji}}=\Delta S_{i\to j},
\label{LDB}  
\end{align}
stating that the log-ratio of forward and backward transition rates
corresponds to the entropy flow along the transition. We note that
time-scale separation is certainly necessary but \emph{not} sufficient for local detailed balance
to emerge; 
a coarse-grained dynamics that
is accurately represented by a Markov-jump process obeying
local detailed balance should, as a rule, be understood as
\emph{not} emerging from lumping, but instead from a reduction
like milestoning \cite{Hartich2021PRX,Hartich2023PRR}.

We remark that the second type of lumping, where we consider some
observable $f(x):\mathbbm{R}^d\to \mathbbm{R}^p$ on the microscopic configuration space
with some $p<d$ generally leads to non-Markovian dynamics for the
reasons listed above. The only situation where Markovian dynamics
emerge upon such type of lumping is when $q_t\equiv f(x_t)$ projects
only onto the $p$ slowest modes of $x_t$ and is essentially orthogonal
to faster-decaying modes (see, e.g., Ref.~\cite{Lapolla2019FP}). Note that
this also includes milestoning onto discrete states localized at metastable cores, which are
actually linear combinations of the slowest modes
\cite{Moro_1995}. Another situation where Markovian dynamics emerge for the
same reasons are projections onto hydrodynamic modes (see, 
e.g., Refs.~\cite{Zwanzig_Hdyn,Zwanzig_A,Spohn_1986,Spohn_1991}). Essentially
all other lumpings lead to non-Markovian dynamics. 

\subsection{Non-Markovian coarse-grained dynamics}
As soon as we include fast motions in the coarse graining via
``lumping'', consider smoothened observations with finite resolution \cite{Dieball2022PRL,Dieball2022PRR},
or coarse grain dynamics without a time-scale
separation, the resulting dynamics will inevitably be
non-Markovian \cite{Mori1965PTP,Zwanzig1973JSP,Zwanzig2001,Netz,Aron2010JSMTE,Dmery2014NJP,Meyer2017JCP}. 
The presence of pronounced memory is not only expected, nowadays several
conclusive methods are available to unambiguously detect \cite{Dima_test} and
quantify \cite{Lapolla2021PRR,Vollmar2024JPAMT} memory in dynamics
observed at coarse resolution.  

We now come to the main part of this perspective---the discussion of
entropy production given coarse grained dynamics with memory.
The main challenges are: Given a coarse-grained observation, what can we
actually deduce about the entropy production in the full, microscopy
system? 

Before delving into specific strategies, we isolate two main points:
(1) How can we \emph{reliably} deduce that the underlying dynamics are
out of equilibrium, i.e., whether $\DSt>0$? (2) Can we infer a lower
bound $\DSb>0$ such that we know $\DSt\ge\DSb$? Given a coarse grained
observation, one can generally of course \emph{not} recover with
certainty the full entropy production or an upper bound, since one may
always miss out on parts of the system that are projected out.    

Moreover, one also has to keep in mind that the total thermodynamic
entropy production $\DSt$ only regards the entropy production that
directly couples to coordinates $x_\tau$, even though there will
generally (i.e., in realistic scenarios extending beyond the
assumptions of the framework of stochastic thermodynamics) be other
sources of entropy production, e.g., thermodynamic cost of maintaining
hydrodynamic flows in the medium entering $a(x)$ in Eq.~\eqref{SDE}. 

The emergence
of memory poses a challenge for thermodynamic inference, since unlike for Markovian positions
(even under time reversal) or velocities (odd under time reversal) it
is generally \emph{not} known how to treat this memory under time
reversal. As mentioned before, memory
effects are manifested as some kind of (anti-)persistence of motion
and in some sense seem to possess some ``oddness'' in time reversal 
but one cannot simply ``flip the sign'' of memory effects.

\subsubsection{Lumped observables of continuous-space dynamics}\label{sec:lump}
In this section we assume that the coarse grained trajectories
$(q_\tau)_{0\le\tau\le t}$ arise from $(x_\tau)_{0\le\tau\le t}$ in Eq.~\eqref{SDE} as a
function $q_\tau=f(x_\tau)$ with $f(x):\mathbbm{R}^d\to
\mathbbm{R}^p$ for some $p\le d$. Alternatively, one might also consider
an intermediate step where $(x_\tau)_{0\le\tau\le t}$ is a Markov-jump
process as discussed in Sec.~\ref{MarkovJ}.
We neglect the possibility of explicit
time-dependence in the drift, diffusion, or jump rates. We also
neglect underdamped motion, although these could be treated similarly
as long as it is known which coordinates represent velocities, and the
function $f$ does not mix position and velocity degrees of freedoms.

 \textbf{(1) Inferring time-reversal asymmetry.} 
Recall from Sec.~\ref{equival} that under the above assumptions
$\DSi=\DSt$ on the level of $\x_\tau$ (denote now by
$\DSi(\x)=\DSt(\x)$). Therefore, if $\DSi(\x)=0$, i.e., the measure
of $\traj$ is symmetric forward and backwards in time, then for
$q_\tau=f(\x_\tau)$ the path measure of $\trajCG$ is also symmetric
in time, i.e., $\DSi=0$ on the level of $q$, denoted by
$\DSi(q)=0$.
There are two ways to see this. The first one is to use that for $\DSi(x)=0$ we have symmetry for all $n$-point densities, $n\in\N$, 
\begin{align}
    &p(\x_{t_1},t_1; \x_{t_2},t_2; \dots; \x_{t_n},t_n)\nonumber\\
    &= p(\x_{t_1},t-t_1; \x_{t_2},t-t_2; \dots; \x_{t_n},t-t_n)\,,\label{npoint x}
\end{align}
which implies 
\begin{align}
    &p(q_{t_1},t_1; q_{t_2},t_2; \dots; q_{t_n},t_n) \nonumber\\
    &= p(q_{t_1},t-t_1; q_{t_2},t-t_2; \dots; q_{t_n},t-t_n)\,,
\end{align}
by integrating on both sides of Eq.~\eqref{npoint x}  over all $\x_{t_i}$ that are mapped to the same value $q_{t_i}$. This may also be understood as in the notation (where expectation runs over all $x_{t_i}$)
\begin{align}
    &p(q_{t_1},t_1; q_{t_2},t_2; \dots; q_{t_n},t_n) 
    = \Elr{\prod_{i=1}^n\delta[q_{t_i}-f(\x_{t_i})]}\nonumber\\
    &\overset{\rm Eq.~\eqref{npoint x}}= \Elr{\prod_{i=1}^n\delta[q_{t_i}-f(\x_{t-t_i})]}\nonumber\\
    &= p(q_{t_1},t-t_1; q_{t_2},t-t_2; \dots; q_{t_n},t-t_n)\,.
\end{align}
Supposing we may take the limit $n\to\infty$ here, this implies $\DSi(q)=0$.
The second way, especially suited when $x_\tau$ is a Markov-jump
process, is via the log-sum inequality, stating that for $(p_i), (p'_i)\ge0$ with $\sum_i p_i=p$, $\sum_i p'_i=p'$, we have $\sum_i p_i\log(p_i/p'_i)\ge p\log(p/p')$. Then grouping all $\x$ that are mapped to the same $q=f(\x)$, and approaching the path measure, we obtain that $\DSi(\x)\ge\DSi(q)$ such that $\DSi(\x)=0$ indeed implies $\DSi(q)=0$ (since by definition $\DSi(q)\ge 0$).
 
This implies that in order to show that $\DSi(\x)>0$ (and hence
$\DSt(\x)>0$) it suffices to show that $\DSt(q)>0$. This is a great
achievement since $\DSt(q)$ is directly accessible from the observed
dynamics. Under the further assumption of steady-state dynamics, we
know that the correct time reversal for overdamped Markov dynamics
$\traj$ is given by simply reading trajectories backwards,
i.e,\ $(\theta x_\tau)_{0\le\tau\le t}=(x_{t-\tau})_{0\le\tau\le t}$. 
Then, since the coarse graining $q_\tau=f(\x_\tau)$ considered here
commutes with time reversal, we deduce that we can infer
$\DSt(q)>0$ simply from antisymmetric correlation functions, or by
finding a time-antisymmetric observable with non-zero mean. 

Such a strategy with antisymmetric correlation functions is the basis
of, e.g., the \emph{variance sum rule} \cite{DiTerlizzi2024S} where the
$\DSt(x)$ is expressed in terms of $\E{x_tF(x_0)-x_0F(x_t)}$, $F(x)$
being the force acting on a particle. Clearly, under the above
assumptions, $\E{q_tf(q_0)-q_0f(q_t)}=0$ in equilibrium (where
$\DSi(x)=0$) for any function $f$, since this observable is antisymmetric
under time reversal. 

This idea directly generalizes to further time-antisymmetric
observables \cite{Vulpiani_review2024arxiv}, such as for example 
$\E{g(q_t-q_0)}$ for any function 
$g(-q)=-g(q)$, e.g.,  $g(q)=\tanh(kq)$ for any length scale $1/k$, or
similarly using $\arctan,\ \sin,\ \sinh,$ etc. One may even add up
information over distinct time-differences \cite{DiTerlizzi2024S} or
over different length scales, e.g., by considering $\int_0^t\E{q_\tau
  f(q_0)-q_0f(q_\tau)}\rmd\tau$ or $\int\rmd
k\E{\tanh[k(q_t-q_0)]}^2$. Further useful observables could be
$\E{(q_t-q_0)q_tq_0}=\E{q_t^2q_0-q_tq_0^2}$ \cite{Pomeau1982JdP,Josserand2016JSP} [or a dimensionless
  version like $\E{(q_t-q_0)q_tq_0}/\var_{\rm s}(q_\tau)^{3/2}$], or
$\E{(q_t-q_0)^3}$ or $\E{(q_t-q_0)(q_t+q_0)^2}$. This old idea was recently \oldblue{rediscovered using the peculiar
antisymmetric observable} $\lim_{t\to\infty}\E{\mathbbm 1(\abs{q_0-q_{-\tau}}>l)(q_0+q_\tau-2q_t)/(q_0-q_{-\tau})}/2$ \oldblue{for some cutoff length $l>0$, where this observable plus $1/2$} was called ``mean back relaxation'' \cite{Knotz2024PRE,Muenker2024NM}.

If any of these observables is non-zero, this implies $\DSi(q)>0$ and
under the given assumptions therefore also $\DSt(\x)>0$, i.e., it
implies, as desired, that the full dynamics are not at equilibrium. Of
course, one always has to be careful when making deductions about the
ensemble average from finite statistics, i.e., one has to make sure
that the given averages \emph{truly} deviate a statistically significant amount
from $0$. 

Other important approaches to deduce non-equilibrium from projected observables include a direct machine
learning approach \cite{Seif2020NP}, or checking violations of manifestly
equilibrium properties such as the transition-path-time symmetry 
\cite{Berezhkovskii_2019,Berezhkovskii2020JPCL,Gladrow2019NC} or
details of the memory \cite{Xizhu2024PRL}. \oldblue{Note that \emph{all} mentioned approaches could be called \textit{non-invasive techniques} or \textit{passive measurements} as they evaluate recorded trajectories instead of, e.g., actively perturbing the system to measure responses.}

\textbf{(2) Lower bounds on total entropy production.} Here we not
only want to know whether the underlying dynamics of a system observed
at coarse resolution is out of equilibrium, but we
also ask \emph{how far} from equilibrium the system operates. This is
a currently a very active field of research. The most important results on lower
bounding entropy production include thermodynamic uncertainty
relations \cite{Barato2015PRL,Seifert2019ARCMP,Koyuk2019PRL,Koyuk2020PRL,Dechant2021PRX,Dieball2023PRL,Kwon2022NJP,arxiv_Crutchfield_TUR},
speed limits and transport bounds
\cite{CSL_3,CSL_7,VanVu2023PRX,Dechant2018PRE,Leighton2022PRL,Dieball2024PRL},
correlation bounds \cite{Dechant2023PRL,Ohga2023PRL,HasegawaPRL2024}, and
thermodynamic inference from hidden Markov models \cite{Skinner2021PRL,Skinner2021PNASU} or snippets \cite{vanderMeer2023PRL,Degunther2024PNASU}.

We emphasize that these results are directly formulated for functions
of Markov dynamics $q_t=f(x_t)$, and since they only require observed
trajectories $\trajCG$ (and do \emph{not} require knowledge of $f$),
they provide lower bounds $0<\DSb<\DSt(\x)$ directly accessible from
experiments. This \oldblue{tacitly} circumvents the problem of time reversal in the
presence of memory in the sense that one no longer needs to revert
time, but instead can simply resort to these inequalities directly. \oldblue{Note that analogous strategies and results are being developed also for underdamped dynamics \cite{Kwon2022NJP,Dieball2024PRL,arxiv_Crutchfield_TUR}. }

Moreover, if $\DSi(q)$ can be evaluated or bounded directly (see,
e.g., Refs.~\cite{Roldan2010PRL,Martinez2019NC,vanderMeer2023PRL,Degunther2024PNASU,Tassilo})
then this also gives a lower bound for $\DSt(\x)=\DSi(\x)$, since as a
consequence of the log-sum inequality we have $\DSi(q)\le\DSi(\x)$
\cite{Esposito2012PRE}. When the lumped process evolves on a discrete
state space further progress can be made in the direction of
the theory of semi-Markov processes
\cite{Martinez2019NC,Tassilo}. A \emph{semi-Markov processes of order $k$} is
a process for which the instantaneous state $q_t$ and the waiting time in said state depend
on the previous $k$ states \oldblue{(but not on the times spent there)}. Thus, $k=1$ corresponds to a renewal
process \cite{Feller1968,Landman_Montroll_Shlesinger1977PNASU} and, if in addition all waiting times are exponentially
distributed, to a Markov process. 

Disregarding waiting-time contributions
(for reasons that will be detailed below) it is possible to estimate
the steady-state entropy production $\dot{S}_{\rm tot}=\frac{\rmd}{\rmd t}\DSt(x)$ from
path measures of $\trajCG$. \blue{If we let $\hat{\gamma}^{(k_i:k_j)}_k$
denote a subsequence $k_i,k_{i+1},\ldots k_j$ of a particular
set of $k+1$ consecutive observed
states with final state $\hat{\gamma}^{({\rm f})}_k$ for $k\ge 1$, and $\mathbbm{P}[\hat{\gamma}^{(k_i:k_j)}_k]$ the probability to observe this (sub)sequence along a (formally infinitely long) trajectory, then  the $k$-th order
estimator is given by \cite{Tassilo} 
\begin{align}
\dot{S}^{\rm est}_k\equiv
\frac{1}{T}\sum_{\hat{\gamma}_k}\mathbbm{P}[\hat{\gamma}^{(1:k+1)}_k]\ln\frac{\mathbbm{P}[\hat{\gamma}^{(\rm
      f)}_k|\hat{\gamma}^{(1:k)}_k]}{\mathbbm{P}[(\theta\hat{\gamma}_k)^{(\rm
      f)}|(\theta\hat{\gamma}_k)^{(1:k)}]},
\label{estimator}
\end{align}
where $T$ is the average waiting time per state in the observed
trajectories
and $\theta\hat{\gamma}_k$ is the corresponding time-reversed sequence of lumped
$k+1$ states.} If $\trajCG$ is a $n$-th order semi-Markov process then
$\dS^{\rm est}_{n+k}=\dS^{\rm est}_{n},\,\forall k>0$ \blue{and
  $\dS^{\rm est}_{n}\le \dot{S}_{\rm tot}$ \cite{Tassilo}}. If
$\hGamma_t=\Gamma_t$ is a Markov process, we have  $\dS^{\rm
  est}_{1}=\dS$. 
Note, however, that when $P(\hat{\gamma}_k)$ are
inferred from trajectory data, undersampling may cause deviations from
these relations.

A particular lower bound, which essentially corresponds to $\dS^{\rm
  est}_{1}$  was used for thermodynamic inference
from non-Markovian lumped data \cite{Yu2021PRL}. However, beyond correctly inferring a
lower bound on $\dS$, the authors also interpreted the
dependence of $\dS^{\rm  est}_{1}$ on the lumping scale, which we
stress is \emph{not} necessarily meaningful and may lead (and in fact
does lead) to misinterpretations (see counterexamples in Ref.~\cite{Tassilo}).

It is important to stress that not all of the above lower bounds rely on
the requirement that the path measure $\mathbbm{P}[\trajCG]$ displays
time asymmetry in the sense that $\DSi(q)>0$ (or $\dot{S}^{\rm
  est}_k>0$). In Ref.~\cite{Dechant2023PRL}, it was recently
demonstrated that one may
infer a positive lower bound on $\DSt(\x)$ from observing $q$
where $\DSi(q)=0$, i.e., to infer a bound on the distance from
equilibrium from a virtually time-reversal symmetric lumped dynamics
$\trajCG$. Note, that proving the mere existence of non-equilibrium in
such virtually time-reversal symmetric lumped dynamics is also
possible via an asymmetry of transition-path times
\cite{Berezhkovskii_2019,Berezhkovskii2020JPCL,Gladrow2019NC}. Further
important progress was made in the direction of transition-based
coarse-graining \cite{Harunari2022PRX,vanderMeerPRX2022}.

While there still remain many challenges, e.g., related to finding optimal lower bounds or experimental
challenges, the overall picture seems to be relatively well understood
as long as we can make sure that time-reversal and coarse graining
commute. In the next section we will show that this latter question,
and its ramifications for inferring dissipation, is in reality far more
difficult. 

\subsubsection{Milestoned observables}

\textbf{Milestoning and \oldblue{time-reversal} do not commute.} 
We now address entropy production and time-reversal symmetry under
milestoning. 
Even though milestoning is clearly necessary to obtain (or at least
to approximate) Markov dynamics (see discussion in Sec.~\ref{MarkovJ} including Figs.~\ref{fig:milestone_2d} and \ref{fig:Kramers}),
and it was shown,
under certain condition, to improve \cite{Blom2024PNASU} or even yield \emph{exact} kinetics and
thermodynamics \cite{Hartich2021PRX,Hartich2023PRR},
it has received much less attention than the naive lumping and the
functions of the dynamics considered before.
We now summarize some of the known results and formulate some open
questions regarding the inference of entropy production from
milestoned dynamics $\trajCG$.

First, note that, by definition, any milestoned trajectory lives
on a discrete state space. Thus, trajectories can be exactly described
by the sequence of states $q_1, q_2,\dots$ and the amount of time
$t_1, t_2, \dots$ spent in the respective states. However, as
generally the case for coarse grained dynamics, even if we decide to
assume $\traj$ to be Markovian, the coarse grained dynamics can have
arbitrarily complicated memory and non-exponential and
history-dependent waiting times. In order to see why milestoning with
``good milestones'' is beneficial, we recall the reasons for
emergence of strong memory effects. First, by projecting onto
localized cores\oldblue{, see Fig.~\ref{fig:milestone_2d},} we inherently eliminate fast local recrossing
events, and focus on long-time/larger-scale, yielding waiting times that
are closer to being exponentially distributed. Second, if we select milestones to be small enough, we
minimize the correlations between entry and exit points of milestones,
and thereby reduce correlations in the sequence of visited states as
well as correlations between waiting times in a given state and the
sequence of preceding states. What is a ``good'' choice of milestones
is a tricky question and in full generality is still an unsolved problem
(see, e.g., Refs.~\cite{Suarez2021JCTC,Schuette2011JCP}). A ``good'' choice of milestones that are to yield a
dynamics as close as possible to Markov will be
the one for which microscopic trajectories will spend only short
periods of time outside any milestone and correlations between
sequences of milestones will be minimal \cite{Hartich2021PRX}. \blue{Note that the (somewhat counterintuitive) concurrent need for small milestones and short times spent outside the milestones for the emergence of Markovian dynamics highlights that approaching such a Markovian limit does by no means exist in general, but instead is a very special case.}
 
For lumped dynamics, we saw before that it is possible to infer lower bounds on the entropy production $\DSt(\x)$ by estimating $\DSi(q)$. In particular, this could be done by separating $\DSi(q)$ into a part obtained from the sequence of states and a part obtained from the times spent in each state before a transition into a certain direction \cite{Martinez2019NC}. Based on this rewriting, however, 
one can quite easily construct examples that\oldblue{, when applying the naive time reversal for milestoned dynamics $\trajCG$,} have $\DSi(q)>0$ while $\DSi(\x)=\DSt(\x)=0$ \oldblue{(note that, unlike for the lumped dynamics, the log-sum inequality is no longer helpful\blue{; for a concrete example of how erroneous entropy production estimates emerge when applying the naive time reversal, see Ref.~\cite{Hartich2021comment}})}.

This, at first surprising, result is caused by the non-commuting of
time reversal and coarse graining, i.e., because for $\trajCG=\mathcal
F[\traj]$ (where $\mathcal F$ denotes the milestoning functional) we
may generally have $\trajCGBW\ne\mathcal F[\trajBW]$ \blue{(note that this statement equally holds for underlying dynamics $\traj$ that represent overdamped or underdamped diffusion or Markov jump processes on discrete state-space)}. We stress that
this result does \emph{not} contradict the results in
\cite{kawai2007dissipation,Gomez}, since a milestoned path weight $\mathbbm{P}(\trajCG)$ is
\emph{not} a marginal path weight of the microscopic process
$\mathbbm{P}(\traj)$ \cite{Hartich2021PRX}.  
The phenomenon
was coined \textit{kinetic hysteresis} and was addressed rigorously
for first-order semi-Markov processes that arise as milestonings
of thermodynamically consistent overdamped diffusions on a graph,
including the Markovian limits \cite{Hartich2021PRX}. Meanwhile, the
results were extended to milestoning of Markov-jump dynamics \cite{Hartich2023PRR}
and milestonings of non-Markovian lumped dynamics \cite{Blom2024PNASU}.

To make this explicit, the waiting-time contribution to entropy
production emerging from a naive time-reversal suggests that any
unequally distributed directional waiting times to exit a given state,
$\psi_{k|j}(t)$ versus $\psi_{j|i}(t)$ for any fixed state $j$ and some states $i,k\ne j$
implies \oldblue{implies $\DSi(q)>0$, even though one could still have $\DSi(\x)=\DSt(\x)=0$, see, e.g., Ref.~\cite{Hartich2021comment}}. Milestoning generally yields non-instantaneous
transition-path times (duration of successful transitions) and in turn
such unequal waiting times, regardless of whether the microscopic
dynamics obeys detailed balance or not
\cite{Wang_2007,Hartich2021PRX,Hartich2023PRR,Hartich2021comment,Blom2024PNASU}. In
other words, in the presence of memory \oldblue{induced by a milestoning procedure} we generally need to
distinguish between mathematical irreversibility and detailed balance
\cite{Wang_2007} and in such scenarios the naive time-reversal does
\emph{not} correspond to physical time-reversal.

The fact that there are examples with $\DSi(q)>0$ that have
$\DSi(\x)=\DSt(\x)=0$ \cite{Hartich2021comment,Blom2024PNASU},  where
in both cases, the backwards dynamics is naively defined to be the
dynamics read backwards, has far-reaching consequences. Whenever the
presence of milestoning cannot be excluded 
\blue{(in particular, in all theoretical works claiming to address a general class of experiments)}, even if 
we are allowed to assume that the microscopic dynamics $\traj$ obeys steady-state dynamics as
in Eqs.~\eqref{SDE} or in Sec.~\ref{MarkovJ}, we cannot bound the
entropy production of $\x$ by the time-reversal asymmetry of
$\trajCG$, i.e., by assessing directly $\DSi(q)$. Therefore, all
aforementioned 
inference strategies are prone to fail for milestoned processes, i.e., one can construct
counterexamples (see Ref.~\cite{Hartich2021comment}) to the \oldblue{inference strategies cited in Sec.~\ref{sec:lump}}. Moreover,
one can no longer deduce non-equilibrium from asymmetries in
correlation functions of based on the observations $\traj$, and approaches like thermodynamic
uncertainty relations and speed limits do not necessarily apply,
although future work might possibly extend these results to milestoned
dynamics.

\textbf{Why should we care?} It is obvious that issues with
a physically consistent time-reversal (i.e., issues with kinetic
hysteresis \cite{Hartich2021PRX}) in the presence of memory arise 
only for milestoned dynamics (including post-lumped milestoned dynamics \cite{Blom2024PNASU}), so it is natural to ask if this is
really an issue that we should care about. The answer is yes, we must, for the following
reasons. First, essentially all experiments have a limited resolution
and frequently detectors  have ``blind spots'' or the signal intensity
is modulated in the detection process and potentially thresholded (e.g., point spread functions of
imaging systems), so the signal we record is in some sense already
``milestoned''.  Second---and this point seems to be frequently
overlooked---whenever we are coarse graining (either lumping or
milestoning) discrete-state Markov jump processes, we are (recall
conditions required for the emergence of Markov-jump processes from
continuous-space dynamics Sec.~\ref{MarkovJ}) implicitly
coarse-graining milestoned dynamics \cite{Hartich2023PRR}. These issues are not merely of
conceptual nature, as there seem to be \emph{no} method available to test
whether a discretely observed non-Markovian processes displays kinetic
hysteresis. \oldblue{Finally, milestoning trajectories (microscopic or lumped) may be useful for thermodynamic inference \cite{Blom2024PNASU}.}

\textbf{How to estimate entropy production for milestoned dynamics?} 
If we have further knowledge about the underlying microscopic dynamics,
such as the existence of some \emph{Markov states} or Markovian
transitions, we may be able to make strict statements about the
entropy production of the underlying dynamics
\cite{Hartich2021PRX,vanderMeerPRX2022,Harunari2022PRX}. 
\blue{Useful information abut the microscopic dissipation is also encoded in transition-path times, i.e.\ the duration of successful transition, corresponding to the timespan between the last exit of the microscopic trajectory from a milestone to the first entrance into a new one. If we indeed} have access to these transition times, then
we can infer
non-equilibrium from violations of transition-path symmetries \cite{Berezhkovskii_2019,Berezhkovskii2020JPCL,Gladrow2019NC}.
Without extra knowledge, but still assuming that $\x$ is a steady-state
overdamped Markov process, the presumably only (bulletproof)
thing to do in general is to 
use saturated-order estimators in
Eq.~\eqref{estimator} \cite{Tassilo} and intentionally disregard waiting times
in order to avoid inconsistencies as
shown in \cite{Hartich2021PRX,Blom2024PNASU,Hartich2021comment}.  

\section{Outlook}
Recently, there has been a surge of interest in understanding and inferring irreversibility of coarse-grained observables. However, despite all efforts and advances, the picture is far from complete and many fundamental (and occasionally elementary) questions remain elusive.\\
\indent\textbf{Semi-Markov processes and milestoning.} First, while the importance of semi-Markov processes (of order $n$) in stochastic thermodynamics is visibly growing, it seems to be underappreciated that parameterizing these requires much more effort and statistics. For example, whereas a Markov process is fully specified with all transition rates between states, a first-order semi-Markov process (i.e.\ renewal dynamics) requires \emph{for each given state} the respective splitting probabilities and waiting-time distributions to transition into each of the neighboring states, respectively (and the waiting-time distributions require a parameterization themselves, say a sum of exponential functions). What is worse, a second-order semi-Markov process requires \emph{for each given state} the splitting probabilities and waiting-time distributions to transition into each of the neighboring states \emph{conditioned on all incoming  transitions}, (and the waiting-time distributions require a parameterization themselves). Typically, we do not know a priori the order of the semi-Markov process unless we know both, the microscopic dynamics as well as the coarse graining. 
Moreover, the parameterization of semi-Markov processes becomes exponentially more challenging with the semi-Markov order. From a practical point of view one should therefore strive to \oldblue{approach} renewal dynamics and hence invoke milestoning. However, formally milestoning has \emph{not} yet been introduced to stochastic thermodynamics beyond renewal processes \cite{Hartich2021PRX}.
Moreover, we lack (sharp) thermodynamic inequalities, such as thermodynamic uncertainty relations and transport bounds, for milestoned dynamics. \blue{In particular, it would be interesting to also consider underdamped underlying dynamics.} Another intriguing question, inspired by Ref.~\onlinecite{Blom2024PNASU}, concerns thermodynamic inference via milestoning of milestoned trajectories.

\textbf{Thermodynamics of semi-Markov processes.} Continuing with semi-Markov processes, rigorous results are only available for renewal dynamics. The results for the thermodynamic entropy production rate for general $n\ge 2$ semi-Markov processes was never proved rigorously. Presumably related, thermodynamic inequalities for semi-Markov processes remain elusive \blue{(with notable exceptions for the special case of first-order semi-Markov processes, see Ref.~\cite{Ertel2022PRE}}).

\textbf{Lost (in) assumption?} Too frequently one does \emph{not} separate between useful and necessary assumptions on the underlying microscopic dynamics \emph{and} the coarse graining, or ignores (or omits) the necessary assumptions all together. {It also happens that crucial assumptions are declared to be ``mild'' even though their validity cannot be tested (e.g., the absence of kinetic hysteresis \cite{Cisneros_2023}). With this in mind, \oldblue{the frequently encountered titles involving} claims of ``general'' or ``universal'' \oldblue{results} seem somewhat inappropriate and as a community we should be more faithful and self-critical, and strive to finding counterexamples to our own statements before declaring ``general validity''. 
\oldblue{Instead, we should attempt to make restricting assumptions in order to make progress while retaining physical and mathematical consistency.}
\\
\indent An example of a useful and plausible (but \emph{not} necessary) assumption is, e.g., the existence of so-called Markovian states \cite{vanderMeer2023PRL,Degunther2024PNASU} in the coarsely-observed dynamics, which allows for substantial progress in thermodynamic inference. Nevertheless, it seems that its validity is not trivial to test unambiguously. Conversely, as explained above, assuming (effectively, i.e., on longest time-scales) Markov-jump dynamics and the absence of kinetic hysteresis may be convenient (in some cases necessary) but is neither plausible nor consistent, and the validity cannot even be tested unless one knows both, the microscopic dynamics \emph{and} the precise coarse graining. Typically we know neither.

\textbf{Active systems.} A particularly challenging topic, even in the absence of any coarse-graining, are active-matter systems \cite{Fodor2016PRL} where it does \emph{not} even seem to be clear how to uniquely/consistently define entropy production given a fully resolved dynamics in a general setting \cite{RoBon2024} (see also Refs.~\cite{Pietzonka2017JPAMT,Padmanabha2023PRE,Fritz2023JSM}). Introducing coarse observations on such systems poses great technical and conceptual challenges, and only little is known and understood at the moment.

The open questions are expected to keep the field engaged for many years to come.

\textbf{Acknowledgments.---}Financial support from the European Research Council (ERC) under the European Union’s Horizon Europe research and
         innovation program (grant agreement No 101086182 to AG) is
         gratefully acknowledged.

\appendix

\blue{
\section{The ``Stratonovich term'' is \emph{not} a Stratonovich term}}
\blue{
As mentioned in the main part, the last term in Eq.~\eqref{action} is often mentioned to be a consequence of the ``Stratonovich discretization scheme'', see, e.g., Ref.~\cite{Lau2007PRE}. We stress that there is \emph{no} such choice of convention, as the dynamics is already specified by the Langevin equation in Eq.~\eqref{SDE} (which for additive noise does not even depend on the convention). In particular, since the dynamics are specified by Eq.~\eqref{SDE}, it should be clear that the action should be unique, which is why Eq.~\eqref{action} is the \textit{only} correct result for the action. The last term only (and obviously) disappears in the small-noise or low-temperature limit $D\to 0$, as considered in Freidlin-Wentzell theory \cite{Freidlin_2012}.

To emphasize the uniqueness of the result for the action, we now list different ways to arrive at Eq.~\eqref{action} which all (if performed correctly) give the same result. Moreover, recall that they also have to give the same result for the action in Eq.~\eqref{action} since the statistics of the paths generated by Eq.~\eqref{SDE} are the same for all conventions, because we only consider additive noise. 

First, note that a mathematical proof for Eq.~\eqref{action} using the notion of smooth tubes to make sense of $\dot x_\tau$ can be found in Theorem 9.1 and Corollary in Chapter VI of Ref.~\cite{Ikeda1981}. Other proofs performed in continuous time derive the action as a Jacobian determinant when transforming the Wiener measure (path measure of the Wiener process $(W_\tau)_{0\le\tau\le t}$) to the path measure of $\traj$. In such proofs one has to choose a convention for integrals such as $I\equiv \int_{t_1}^{t_2}f(t)\delta(t-t_1)dt$, where the value $I=f(t_1)/2$ for this integral is called ``Stratonovich convention'' while the value $I=0$ would be referred to as It\^o (which would give rise to the action in Eq.~\eqref{action} without the last term). However, there is no freedom to choose here since dealing with functional calculus always requires taking the $1/2$ convention here (see, e.g., Refs.~\cite{Fox1986PRA,Fox1987JSP,Dieball2023JPA}).

Another strategy to derive the action is to discretize the process in time, different discretization schemes \emph{virtually} appear to correspond to different conventions. However, in the limit of continuous time, the different schemes all yield the action in Eq.~\eqref{action}, see Eqs.~(17)-(30) in Ref.~\cite{Gladrow2021PRX}. Note that Ref.~\cite{Gladrow2021PRX} even shows the validity of Eq.~\eqref{action} (and its failure without the last term) quantitatively for experimental data. 

Another reference for Eq.~\eqref{action} is Eq.~(2.2b) in Ref.~\cite{Wiegel_1986}, where also in Eq.~(2.2b) it is stated that the last term only vanishes in the small-noise or low-temperature limit, in agreement with Freidlin-Wentzell theory \cite{Freidlin_2012}.
}

\bibliography{perspective}

\end{document}